\documentclass[reprint,aps,pra,superscriptaddress,longbibliography,
amsmath,amssymb]{revtex4-1}
\usepackage{mathtools}
\usepackage{graphicx}
\usepackage{booktabs,multirow}
\usepackage{braket}
\usepackage{textcomp}
\usepackage{yfonts}
\usepackage{comment}
\usepackage[
  bookmarks=false,
  colorlinks,
  linkcolor=blue,
  urlcolor=blue,
  citecolor=blue,
  plainpages=false,
  pdfpagelabels,
  final,
  breaklinks=true
]{hyperref}

\usepackage{lineno}
\usepackage{soul}

\begin{document}

\graphicspath{{figures/}} 
\allowdisplaybreaks

\newcommand{\ui}{\mathrm{i}}
\newcommand{\ud}{\mathrm{d}}

\newcommand{\bs}{\boldsymbol}
\newcommand{\bt}{\textbf}

\definecolor{mypink}{rgb}{.78, 0., 0.38} 
\newcommand{\rgc}[1]{\textcolor{mypink}{#1}}
\newcommand{\maa}[1]{\textcolor{blue}{#1}}

\title{Closed forms for  spatiotemporal optical vortices and sagittal skyrmionic pulses}

\author{S.~Vo}
\affiliation{The Institute of Optics, University of Rochester, Rochester, NY 14627, USA}
\author{R. Guti\'errez-Cuevas}
\affiliation{Institut Langevin, ESPCI Paris, Université PSL, CNRS, 
75005 Paris, France}
\author{M.~A.~Alonso}
\email{miguel.alonso@fresnel.fr}
\affiliation{Aix Marseille Univ, CNRS, Centrale Marseille, Institut Fresnel, 
UMR 7249, 13397 Marseille Cedex 20, France}
\affiliation{The Institute of Optics, University of Rochester, Rochester, NY 14627, USA}
\affiliation{Laboratory for Laser Energetics, University of 
Rochester, Rochester, NY 14627, USA}

\date{\today}


\begin{abstract} 
	Spatiotemporal optical vortices (STOVs) are short pulses that present a vortex whose axis is perpendicular to the main propagation direction. We present analytic expressions for these pulses that satisfy exactly Maxwell's equation, by applying appropriate differential operators to complex focus pulses with Poisson-like frequency spectrum. We also provide a simple ray picture for understanding the deformation of these pulses under propagation. Finally, we use these solutions to propose a type of pulse with sagittal skyrmionic polarization distribution covering all states of transverse polarization.
\end{abstract}

\maketitle


\section{Introduction}

A broad range of short light pulses with interesting distributions in space and time have been studied theoretically and implemented experimentally. For a description of these pulses, the reader can consult extensive reviews of this topic \cite{hernandez2008localized,turunen2010propagation,yessenov2022space}. 
From the theoretical point of view, only a small subset of these pulses can be expressed as closed form solutions of Maxwell's equations. One such case is that of donut-shaped (or toroidal) pulses that  present rotational symmetry about the main direction of propagation, and include a nodal line along this axis of symmetry \cite{Hellwarth1996,Zdagkas2019Singularities,Zdagkas2022Observation}. 
%
%
Recently, a different kind of donut-shaped pulse, referred to as spatiotemporal optical vortex (STOV), has received significant attention \cite{Sukhorukov2005,Bliokh2021Spatiotemporal,Jhajj2016,Hancock2019,chong2020generation,hancock2021second,huang2021properties,wan2022photonic,bliokh2012spatiotemporal,mazanov2021transverse}. Unlike the toroidal pulses mentioned earlier, STOVs present the peculiarity of carrying a phase vortex whose axis is perpendicular to their main direction of propagation, hence resembling advancing hurricanes or cyclones.
To our knowledge, no closed-form expression valid beyond the paraxial regime has been proposed for STOVs. 
 
The current work has several goals:

The first is to propose a relatively simple closed-form expression for representing STOVs, both for scalar and electromagnetic (vector) fields, which are exact solutions of the wave or Maxwell equations in free space, and are then valid beyond the paraxial regime. 
Using this theoretical model, we find conditions to make these vortices as isotropic as possible at the focal time. 
The key to finding these solutions is the use of the
complex focus model \cite{Berry_1994,Sheppard1998,Sheppard1999} 
which consists on assigning complex spatial coordinates to the focal point of a monochromatic spherical wave. 
This method has been used, for example, to construct bases for describing highly-focused electromagnetic fields \cite{Moore2009Closed,Moore2009Bases,GutierrezCuevas2017Scalar} which, in turn, simplify the description of Mie scattering \cite{moore2008closed,gutierrez2018lorenz}.
The complex focus method can be extended to the study of time-dependent pulsed electromagnetic fields by integrating these monochromatic solutions weighted by a given frequency spectrum \cite{Heyman1986,Heyman1989,heyman2001gaussian,Saari2001Evolution,Lin2006Subcycle,April2010}. 
In particular, a Poisson-like spectrum leads to a closed-form expression and allows identifying the resulting complex fields as the analytic signal representation of the real fields \cite{porras1998ultrashort,Caron1999Free}. 

The second goal is to provide a simple, intuitive explanation to the asymmetric deformation that these pulses experience as they propagate. While these deformations are naturally incorporated into closed-form expressions, they can be better understood by using a simple ray-based model that is compatible with the complex-focus picture.  

The third and final goal of this work is to use these models to propose a new type of electromagnetic pulse that presents over its sagittal plane a ``Stokes-skyrmionic'' distribution of polarization, in which all paraxial states of full polarization are represented. 
This pulse is motivated by recent interest in optical fields presenting skyrmonic structures, which are localized regions where a field property, such as paraxial polarization or the spin of nonparaxial vector electric fields, covers the surface of a sphere \cite{Skyrme,beckley2010full,first_optical_skyrmions,skyrmion_spin_evanescent,skyrmions_Rodrigo_Pisanty,paraxial_skyrmionic_beams,optical_hopfion}. 
However, unlike for the simple case of full Poincar\'e beams \cite{beckley2010full,paraxial_skyrmionic_beams,Guti_rrez_Cuevas_2021} where this structure is present over transverse planes, the pulsed fields described here cover the Poincar\'e sphere over sagittal planes parallel to the propagation axis. 

\section{The fundamental scalar pulsed field}

\begin{figure*}
    \begin{center}
    \includegraphics[width=.9\linewidth]{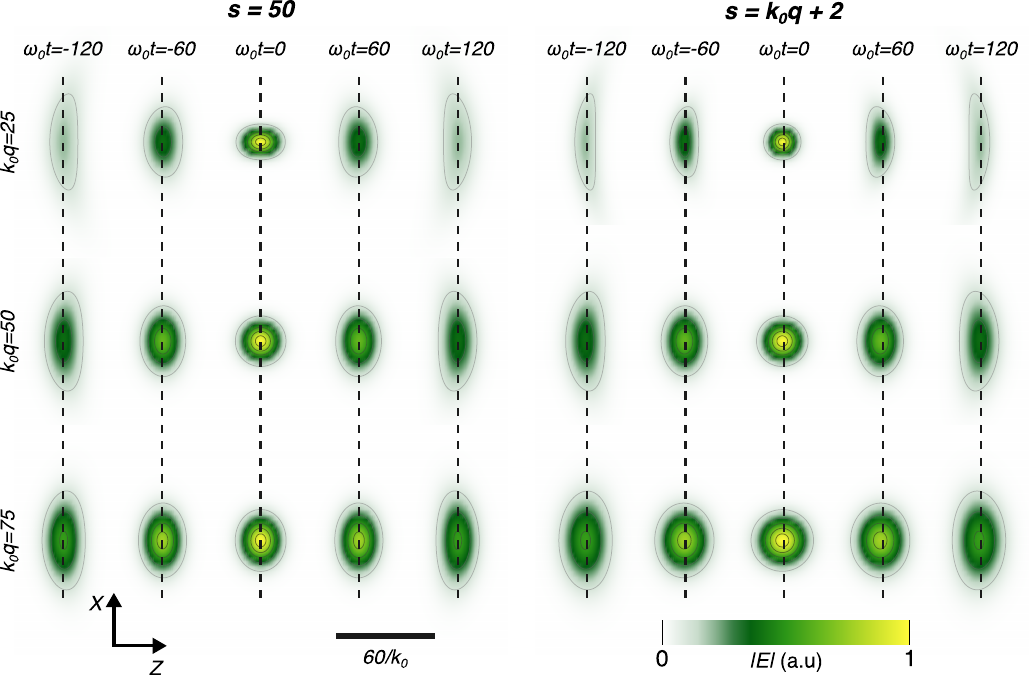}
    \caption{Short-time averaged intensity of a round pulse at five equally spaced times centered at $t=0$ indicated in white in units of $1/\omega_0$, for increasing values of $q$ from top to bottom. 
    For the plots on the left side $s=50$, whereas for the plots on the right side $s=k_0 q +2$.
    }\label{fig:blob}
    \end{center}
\end{figure*}

Monochromatic complex focus fields were proposed as exact closed-form singularity-free solutions to the Helmholtz equation $\nabla^2U+(\omega/c)^2U=0$ (where $\omega$ is the frequency and $c$ is the speed of light in vacuum) that generalize Gaussian beams into the nonparaxial regime \cite{Berry_1994,Sheppard1998,Sheppard1999}. 
They take the form of a standing spherical wave focused at a complex position according to
\begin{align}
U\left(\boldsymbol{r},\omega\right)&= 2 \ui c\,\exp\left(-\frac{\omega |\boldsymbol{q}|}c\right)\, U_0\left(\omega\right)\frac{\sin{\left(\frac{\omega}{c}R\right)}}{\omega R},
\label{eq:complexfield}
\end{align}
where  $U_0\left(\omega\right)$ is an amplitude factor (written here as a function of frequency in anticipation of a derivation that follows) and $R=\left[\left(\boldsymbol{r}-\boldsymbol{\rho_0}\right)\cdot\left(\boldsymbol{r}-\boldsymbol{\rho_0}\right)\right]^{1/2}$ is the complex distance between the observation point $\boldsymbol{r}$ and the complex coordinate $\boldsymbol{\rho_0}=\boldsymbol{r_0}+\ui \boldsymbol{q}$, with $\boldsymbol{r_0}$ being the focal point and $\boldsymbol{q}$ determining the directionality of the field. 
We also include a damping factor $\exp{\left(-\omega |\boldsymbol{q}|/c \right)}$ to guarantee convergence \cite{April2010}.
For a field whose main direction of propagation is aligned with the positive $z$ axis, $\boldsymbol{q}$ can be written as $q \boldsymbol{\hat{z}}$ with $q>0$ and where $\boldsymbol{\hat{z}}$ is the unit vector in the $z$ direction. 
For $\omega q/c\gg 1$, the field tends towards a Gaussian beam whose Rayleigh range is $q$. 
For simplicity we place the focal point $\boldsymbol{r_0}$ at the origin, so that we can write $R=\sqrt{x^2+y^2+(z-\ui q)^2}$.

Pulsed beams with closed-form solutions
can also be constructed by using the concept of complex focus \cite{April2010}. 
The key is to integrate over frequency the product of $\exp(-\ui\omega t)$ and the monochromatic complex focus solution in Eq.~(\ref{eq:complexfield}) with $q$ (the Rayleigh range) being the same for all frequencies.
This condition leads to isodiffracting pulses \cite{Wang1997Space,Caron1999Free,Feng2000Spatiotemporal} and is used to model pulses generated inside cavities where the Rayleigh range is determined by the geometry of the cavity and is independent of the frequency.
By writing $U_0\left(\omega\right)=\omega A(\omega)$, the result is expressible in terms of the inverse Fourier transform of $A(\omega)$.
One simple option would then be to choose $A$ as a Gaussian centered at a frequency $\omega_0$. One minor issue with this choice, though, is that the simple result would involve some amount of negative frequency  contributions. 
This is not a problem if all we require is a closed form for the real field. 
However, sometimes it is convenient to have a form for the complex analytic signal representation of the field, e.g. to calculate easily short-time-average intensities or Poynting vectors. 
It is then better to use a spectrum that allows a simple closed-form solution even when integrating only over positive frequencies. 
A frequency spectrum following a Poisson-like distribution satisfies these properties \cite{Caron1999Free} and therefore will be used:
\begin{align}
U_0(\omega)=\left(\frac{s}{\omega_0}\right)^{s+1}\frac{\omega^{s} \exp\left(-\frac{s\omega}{\omega_0}\right)}{\Gamma(s+1)}\Theta(\omega),
\label{eq:Poissonspectrum}
\end{align}
where $\Theta(\cdot)$ is the Heaviside distribution that ensures that only positive frequencies are involved, $\omega_0$ is the peak frequency, and $s$ controls the width of the spectrum and, thus, the pulse duration.
It is shown in Appendix~\ref{sec:poisson} that the resulting field is simply given by 
\begin{align}
    E\left(\boldsymbol{r},t\right)=\frac{1}{\omega_0 R} \left[\frac{1}{\left(1+ \ui \frac{\omega_0}{s}t_-\right)^s}-\frac{1}{\left(1+ \ui \frac{\omega_0}{s}t_+\right)^s}\right],
    \label{eq:Poissonfield}
\end{align}
where $t_\pm=t\pm R/c-\ui q/c$.
The actual field corresponds to the real part of the complex function in Eq.~\eqref{eq:Poissonfield}. 
However, this complex expression facilitates the computation of what we call the short-time averaged intensity as $|E|^2$, which corresponds approximately to the intensity averaged over an optical cycle.

Figure~\ref{fig:blob} shows the short-time-averaged intensity of the resulting pulse for various values of $q$. 
These plots also show the dependence of the spatial dimensions of the pulse on the parameters $q$ and $s$. 
As $q$ changes for fixed $s$, the longitudinal size of the pulse stays constant while the transverse size varies. 
Therefore, the STOV's width (and level of collimation) is controlled by $q$, while its length (and duration) is determined by $s$. 
As shown in Appendix~\ref{sec:blob}, by considering the short-time averaged intensity for $t=0$, the transverse extension of the pulse is approximately given by the standard expression in terms of the Rayleigh range $q$, namely $\sqrt{q/k_0}$, where $k_0=\omega_0/c$. The longitudinal extension of the pulse, on the other hand, is roughly proportional to $\sqrt{s}/k_0$. The condition for making the pulse round is then found to be given approximately by
\begin{align}
    s
    =k_0 q+2,
    \label{eq:expressions}
\end{align}
valid for $k_0q$ sufficiently greater than unity. 
Note that pulses approaching the paraxial condition have Rayleigh ranges that are much larger than the wavelength for the central frequency, so that the first term dominates, and the second can be neglected.
We do keep the second term of Eq.~(\ref{eq:expressions}) as a first nonparaxial correction.
%
The width of the pulse at $t=0$ in any direction is then given approximately by $\sqrt{q/k_0}$. 
Figure~\ref{fig:blob} shows the short-time-averaged intensity of the resulting round pulse at five times around $t=0$, for increasing values of $q$ from top to bottom.


\section{Scalar STOV}

\begin{figure*}
    \begin{center}
    \includegraphics[width=.9\linewidth]{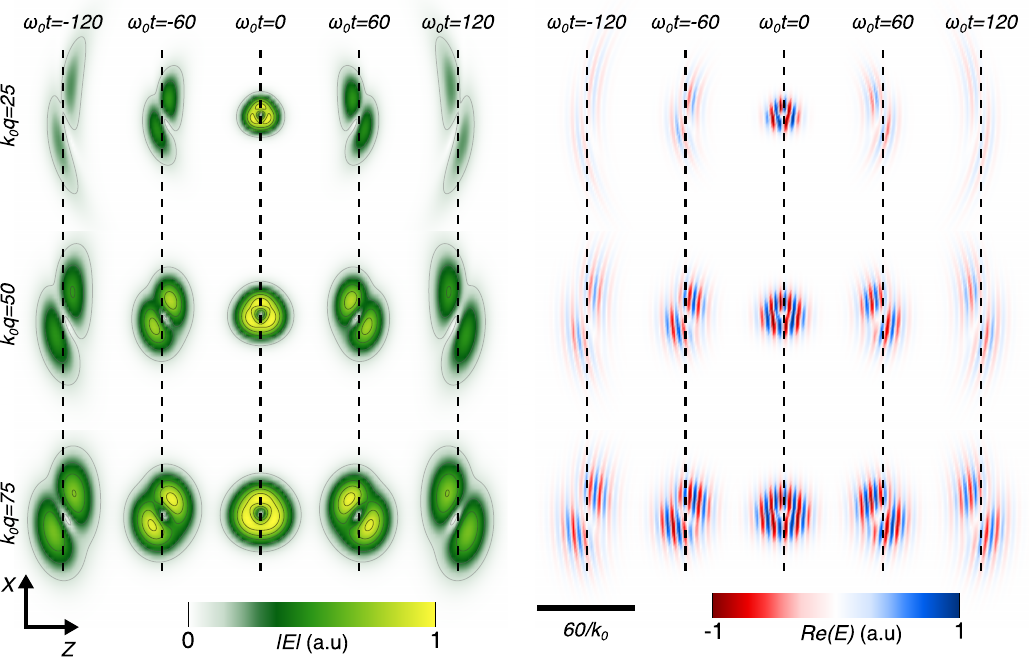}
    \caption{Short-time averaged intensity (left) and real amplitude (right) for scalar STOVs at five equally spaced times centered at $t=0$ indicated in white in units of $1/\omega_0$, 
    for values of $k_0 q= 25, 50, \text{ and } 75$ and $s=k_0q+2$.}
    \label{fig:donut}
    \end{center}
\end{figure*}

The expression for the round blob pulse just described can be turned into one representing a STOV that carries transverse orbital angular momentum (OAM) by applying an appropriate differential operator. 
By choosing the OAM to point in the $y$ direction, the differential operator is constructed as 
\begin{align}
\widehat{W}=\frac{1}{k_0}\partial_x\pm\ui\left(\frac{\alpha}{\omega_0} \partial_t+ \frac{\beta}{k_0}\partial_z+ \ui \gamma \right),
\label{eq:scalaroperator}
\end{align}
where  $\alpha$, $\beta$, and $\gamma$ are constant parameters to be determined. 
The justification for this operator is the following: the derivative in $x$ on its own introduces a nodal plane at $x=0$ given the symmetry of the original pulse, while the combination in parentheses on its own introduces a nodal surface that is normal to the $z$ axis. 
The constant coefficients $\alpha,\beta,\gamma$ can then be chosen so that at $t=0$ the pulse's short-time average intensity is symmetric around $z=0$ and as round as possible. 
These conditions are investigated in Appendix~\ref{sec:stov}, where the following approximate results are found: 
\begin{align}
    \alpha&\approx\frac2{3}-\frac{26}{9s},\qquad
    \beta\approx\frac5{3}-\frac{32}{9s},\qquad
    \gamma\approx-1+\frac{7}{3s}.
    \label{eq:scalarcoefficients}
\end{align}
Note that we included a first nonparaxial correction in each of these coefficients, expressed in terms of the inverse of the dimensionless parameter $s=k_0q+2$ (assumed to be considerably larger than unity). 
The scalar STOV is then simply constructed as
\begin{align}
    E_{\rm STOV}(\boldsymbol{r},t)=\widehat{W}E(\boldsymbol{r},t).
    \label{eq:scalarSTOV}
\end{align}
Figure \ref{fig:donut} shows the short-time averaged intensity and the real amplitude of these scalar STOVs for different values of $q$.
Note that this operator could be applied repeatedly to generate higher order vortices, although this might require the adjustment of the constant coefficients. Also, by including a term including a derivative in $y$, together with an adjustment of the constant coefficients, one could introduce a vortex whose axes are in any desired direction \cite{wan2022photonic}.


\section{Ray picture}
One can understand the deformation of the STOV away from $t=0$ by using a simple ray picture in two dimensions. 
Consider a family of rays that are perfectly focused at an on-axis point $(x,z)=(0,z_0)$ and that  have maximum slope $\tau$. If we parametrize each ray in terms of a parameter $\xi$, the transverse $x$ coordinate of these rays in terms of $z$ can be written as
\begin{align}
    X(z,\xi) =(z-z_0)\,\tau \cos{\xi}=\tau\,\Re\left[\left(z-z_0\right)\exp(\ui\xi)\right].
    \label{eq:raysfocus}
\end{align}
If we now let $z_0$ become purely imaginary by substituting $z_0=\ui q$, this equation becomes
\begin{align}
    X(z,\xi) =\tau\,\Re\left[\left(z-\ui q\right)\exp(\ui\xi)\right]= q\tau  \sin{\xi}+z\,\tau \cos{\xi},
    \label{eq:rays}
\end{align}
This family of rays resembles a lateral projection of a ruled hyperboloid, and presents a hyperbolic caustic composed of two branches with vertices at $(x,z)=(\pm q\tau,0)$. This hyperbolic profile is consistent with the fact that, in the paraxial regime, Gaussian beams and their related modes such as Hermite- and Laguerre-Gaussian beams are associated often with families of rays described by a caustic that expands hyperbolically \cite{Berry2008,AlonsoDennis}, and that in the nonparaxial regime complex focused fields are connected with oblate spheroidal coordinates \cite{Landesman1988Gaussian,Saari2001Evolution}.

\begin{figure}[h]
\begin{center}
\includegraphics[width=.95\linewidth]{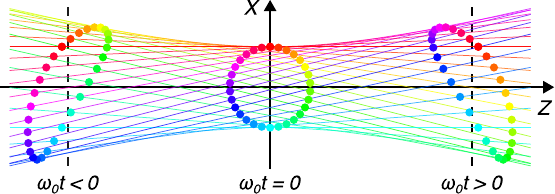}
\caption{Ray-based picture for the deformation of a STOV with propagation. The STOV is represented as a set of particles (color dots) that form a circle at $t=0$. All particles travel at the same speed ($c$), each along a different ray (shown in the same color as the corresponding particle). The picture shows also the deformed configuration of these particles at a later time.}\label{fig:RaysSketch}
\end{center}
\end{figure}

To now mimic the behavior of a STOV, we consider a set of ``particles'', each traveling along a ray at the same speed (the speed of light). 
The initial conditions for these particles are such that they trace at $t=0$ a circle of radius $w$ centered at the origin. 
Note that two rays cross each point along this circle, so care must be taken in assigning which of these rays corresponds to the path of the corresponding particle. 
We can choose, for example, the rays with negative (positive) slope for all the points for which the particle is at $z>0$ ($z<0$) at $t=0$, as shown in Fig.~\ref{fig:RaysSketch}, where the particles and their trajectories are identified by using different colors. 
This situation corresponds to OAM in the positive $y$ direction (into the page), since the evolution of the loop formed by the particles implies a clockwise rotation. 
Notice that, in addition to this rotation and the global displacement to the right, the loop gets deformed with propagation, with the bottom part moving ahead of the top part.
We also observe bunching near the two caustics.

\begin{figure}
\begin{center}
\includegraphics[width=.95\linewidth]{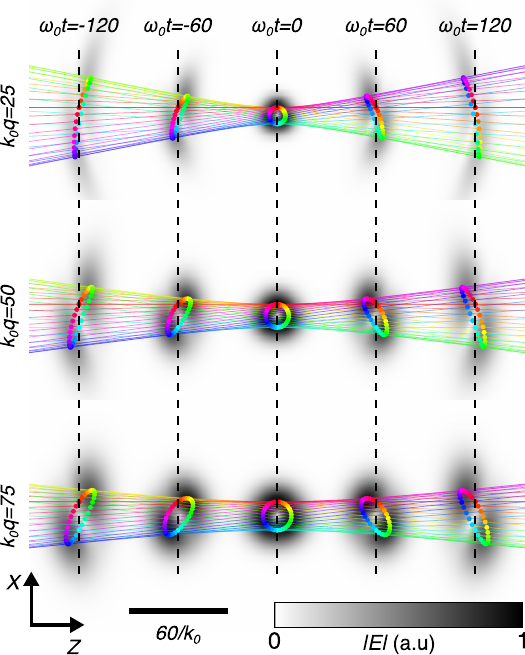}
\caption{
Superposition of ray/particle picture of STOV deformation under propagation to the short-time-averaged intensity of scalar STOVs at several times, and for values of $k_0 q= 25, 50, \text{ and } 75$ and $s=k_0q+2$.
}\label{fig:Rays}
\end{center}
\end{figure}

Figure~\ref{fig:Rays} shows the superimposition of such sets of rays and particles on the short-time averaged intensity plots round STOVs, with 
width $q\tau=\sqrt{q/k_0}$, that is, $\tau=1/\sqrt{k_0q}$. 
Notice that the particle distribution resembles the STOV's intensity maxima, and this correspondence 
improves as we approach the paraxial regime (i.e. $q$ becoming large). Note also that the particle distribution undergoes a sort of rotation with time around its centroid: the particles at the top and bottom of the loop correspond to rays that touch the caustics near the region of the loop. This rotation echoes the orbital angular momentum carried by the STOV.

\section{Electromagnetic STOV}

We now apply similar techniques 
in order to generate closed-form expressions for electromagnetic STOV. 
For this, we apply an operator \cite{Sheppard2000Polarization,Alonso2011} based on the Hertz potential formalism, which transform a scalar pulse into a vector pulse that satisfies both the wave equation and Gauss' divergence condition. 
This operator is given by
\begin{align}
    \widehat{\bf V}_{\boldsymbol{p}}=c^2(\boldsymbol{p}\cdot\nabla)\nabla+c\,\partial_t(\hat{\boldsymbol{z}}\times\boldsymbol{p})\times\nabla-\boldsymbol{p}\,\partial_t^2,
    \label{eq:electromagneticoperator}
\end{align}
where $\boldsymbol{p}$ is a dimensionless constant vector. 
This vector can be real or complex, and its transverse components correspond approximately to the polarization of the pulse near the $z$ axis as it propagates far from the focal zone. For simplicity we then choose $\boldsymbol{p}$ to be constrained to the $xy$ plane. 
For example, choosing $\boldsymbol{p}\propto(1,\pm{\rm i},0)$ leads to STOVs with definite helicity. 
The operator in Eq.~\eqref{eq:electromagneticoperator} commutes with the D'Alembert (wave) operator and its divergence vanishes, so when applied to a scalar solution of the free-space wave equation it gives a vector form of the electric field that is fully consistent with Maxwell's equations. 
It is designed such that, when applied to a scalar pulse traveling in the positive $z$ direction close to the paraxial regime and whose spectrum is not too broadly distributed around a central frequency $\omega_0$, the resulting vector solution resembles the scalar field times $\omega_0^2\boldsymbol{p}$. 

\begin{figure}
\begin{center}
    \includegraphics[width=.95\linewidth]{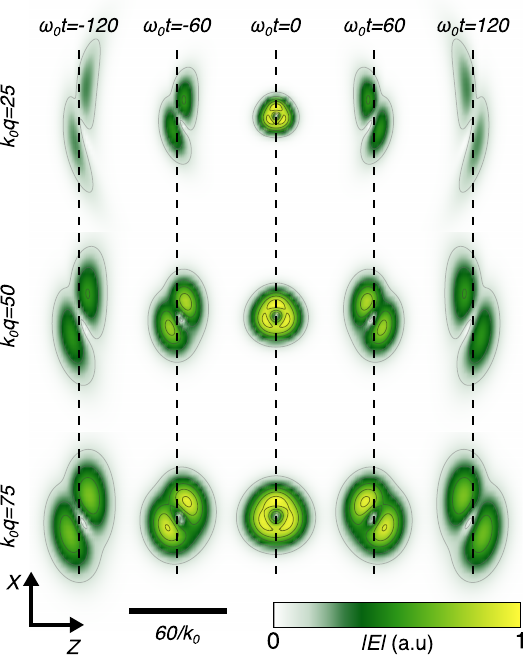}
\caption{Short-time averaged intensity for Electromagnetic STOVs sith definite helicity, at five equally spaced times centered at $t=0$ indicated in white in units of $1/\omega_0$, and for values of $k_0 q= 25, 50, \text{ and } 75$ and $s=k_0q+2$.}
\label{fig:vecdon}
\end{center}
\end{figure}

Closed-form vector solutions for STOVs can be constructed by applying this operator to the scalar STOVs described in the previous section:
\begin{align}
    {\bf E}_{\rm STOV}(\boldsymbol{r},t;\boldsymbol{p})=\frac1{\omega_0^2}\widehat{\bf V}_{\boldsymbol{p}}\widehat{W}E(\boldsymbol{r},t).
\end{align}
The operators $\widehat{\bf V}_{\boldsymbol{p}}$ and $\widehat{W}$ commute, so their order is not important. 
However, the application of the operator $\widehat{\bf V}_{\boldsymbol{p}}$ not only imparts a vectorial character to the field, but also causes some changes to its intensity profile, and hence deforms the STOV. 
As shown in Appendix~\ref{sec:elect}, the approximate rotational symmetry of the 
STOV can be partially restored by adjusting the parameters within $\widehat{W}$ to slightly different values from those that achieve a round scalar STOV:
\begin{align}
    \alpha\approx \frac{6}{7},\qquad \beta\approx \frac{13}{7}-\frac{237}{98 s},\qquad \gamma\approx-1+\frac{237}{98 s}.
    \label{eq:electromagneticcoefficients}
\end{align}
Figure~\ref{fig:vecdon} shows the short-time averaged intensity (the squared norm of the complex electric field) for these electromagnetic STOVs for several values of $s$ from top to bottom and at different times, with $\boldsymbol{p}\propto(1,\ui,0)$. Note that, at the focal time, these pulses have a fairly round profile with a small trigonal deformation. In fact, as shown in Appendix~\ref{sec:elect}, achieving this level of roundness also requires allowing the STOV to shift laterally by an amount $x_0=\mp\lambda_0/14 \pi$. This shift of a fraction of a wavelength can be regarded as a spin-orbit effect, similar to other spin-induced transverse shifts in optical beams \cite{bliokh2010angular,bliokh2013goos–haenchen,bliokh2015spin}. 
%
%

\section{Paraxial limit}

We now study the paraxial limit of the solutions discussed in the previous sections. 
The starting point is the paraxial form of the pulse in Eq.~(\ref{eq:Poissonfield}), which is found in Appendix~\ref{sec:paraxial} to be given by
\begin{align}
  E\left(\boldsymbol{r},t\right)&\approx \frac{1}{z - \ui q} \exp\left\{\ui k_0\left[z-ct + \frac{\rho^2}{2\left(z-\ui q\right)}\right]\right\}\nonumber\\
  &\times\exp\left\{-\frac{k_0}{2q}\left[z-ct + \frac{\rho^2}{2\left(z-\ui q\right)}\right]^2\right\}.
  \label{eq:Gaussian_pulse}
\end{align}
This expression is valid for $s \approx k_0q \gg1$. 
Notice that the first line corresponds to a monochromatic Gaussian beam of frequency $\omega_0$, and the exponential factor in the second line is responsible for the longitudinal localization and the curvature of the intensity profile away from the focal region.

To find the corresponding formula for the scalar STOV we apply the operator in Eq.~\eqref{eq:scalaroperator} to this paraxial pulse, using either the coefficients in Eq.~\eqref{eq:scalarcoefficients} or those in Eq.~\eqref{eq:electromagneticcoefficients}. 
In both cases, the result is proportional to the pulse itself times a factor corresponding to a ratio of polynomials involving powers of $k_0q$. 
By keeping only the leading terms in this factor for $k_0 q\gg1$ we find the same result for both sets of coefficients:
\begin{align}
    E_{\rm STOV}\left(\boldsymbol{r},t\right)\approx\frac{q^2[\pm(ct-z)-\ui x]}{(q+\ui z)^3}
    E\left(\boldsymbol{r},t\right).
\label{eq:Gaussian_donut}
\end{align}


Finally, the corresponding formula for the electromagnetic STOV is found by applying the operator in Eq.~\eqref{eq:electromagneticoperator}. 
Similarly to the scalar STOV, the result can be written as a vector composed of three components being products of the pulse itself times factors also corresponding to ratios of polynomials comprising in particular the components of the constant vector $\boldsymbol{p}$. 
Keeping the leading terms in those factors for $k_0q\gg1$, we find 
 \begin{align}
{\bf E}_{\rm STOV}(\boldsymbol{r},t;\boldsymbol{p}) \approx \frac{2 q^{7} \omega_0^5 \left[\pm(ct-z) + \ui x \right]}{\ui c \left( z -  \ui q \right) ^8} E\left(\boldsymbol{r},t\right) \boldsymbol{p},
\end{align}
where we used the assumption that $\boldsymbol{p}$ was chosen such that $p_z=0$. 
Adding the next leading term in the factors multiplying the original pulse introduces first order corrections comprising, on one hand, an additional $z$-component to the result, and on the other hand an additional contribution to the first two components.

\section{Pulses with sagittal skyrmionic distributions}

\begin{figure}
    \begin{center}
        \includegraphics[width=.95\linewidth]{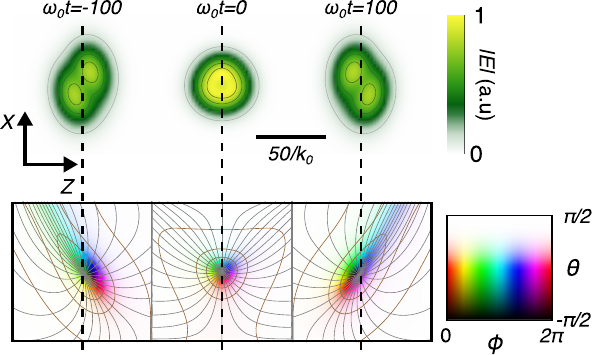}
    \caption{(a,b) Short-time averaged intensity and (c,d) transverse polarization distribution for a pulse presenting a sagittal skyrmionic distribution, at (a,c) $t=0$ and (b,d) $t=40/\omega_0$. In (c,d) the outermost visible $\theta$ contour corresponds to $\theta=-\frac{7 \pi}{16}$}
    \label{fig:skyrm}
    \end{center}
\end{figure}

There has been interest recently in the generation of optical analogs \cite{Poincare_skyrmions,first_optical_skyrmions,skyrmion_spin_evanescent,skyrmions_Rodrigo_Pisanty,optical__merons_PRL,optical_hopfion,optical_plasmonic_merons,Shen_bimerons,shen2022topological,optical__merons_Zhang,paraxial_skyrmionic_beams,Airy_beams_merons,cartography_skyrmions,spin_merons_polygons,F3DP_fields} of skyrmions \cite{Skyrme,magnetic_skyrmions_review,manton1987geometry,esteban1986direct}, which are topological textures in which a spherical parameter space is fully covered in a physical flat (sub)space, according to some rules. 
This includes pulsed solutions where the electric and magnetic fields cover all directions over transverse planes \cite{shen2021supertoroidal}. 
A different type of skyrmionic distribution are the so-called full Poincar\'e beams \cite{beckley2010full}, which are monochromatic paraxial solutions of the wave equation in which all possible states of full polarization are represented in any cross-section of the beam. The spherical parameter space in this case is the Poincar\'e sphere, whose spherical coordinates,
\begin{subequations}
    \begin{align}
    \theta&=\arccos\left[\frac{2{\rm Im}(E_x^*E_y)}{|E_x|^2+|E_y|^2}\right]\in[-\pi/2,\pi/2]\\
    \phi&={\rm arg}\left[|E_x|^2-|E_y|^2+2\ui{\rm Re}(E_x^*E_y)\right]\in[0,2\pi),
\end{align}
\end{subequations}
characterize the ellipticity/handedness and orientation of the polarization ellipse, respectively. The sphere is mapped onto each transverse plane of these beams according to a stereographical map. Nonparaxial versions of these beams have also been defined \cite{Guti_rrez_Cuevas_2021}.

The formulas described in previous sections then suggest that it is possible to create a pulsed distribution of this kind, where polarization is covered not over a transverse plane but over a longitudinal one. The key is to combine an electromagnetic STOV with some choice of $\boldsymbol{p}$ and a copropagating electromagnetic blob with the orthogonal choice of $\boldsymbol{p}$, so that the blob is centered at the STOV's vortex. Here we choose these vectors to give each of these components definite helicity. To make the connection more direct, we consider a pulse close to the paraxial limit ($q=100c/\omega_0$) and we neglect the electric vector's $z$ component in the calculation of the Stokes parameters. The resulting pulse can then be written as
\begin{align}
    {\bf E}_{\rm skyrm}(\boldsymbol{r},t;\boldsymbol{p}_1;\boldsymbol{p}_2)=\left(\cos{\eta}\widehat{\bf V}_{\boldsymbol{p}_1}\widehat{W}+ \sin{\eta} \widehat{\bf V}_{\boldsymbol{p}_2}\right)E(\boldsymbol{r},t),
\end{align}
where $\boldsymbol{p}_1$ and $\boldsymbol{p}_2$ are two mutually orthogonal polarization vectors, and $\eta$ is a parameter that regulates the polarization state distribution across the pulse's $xz$ section.

These solutions have then the property of presenting essentially all transverse polarization states represented within the sagittal plane $y=0$ (as well as over other nearby planes parallel to this one), normal to the polarization plane. This coverage gets deformed as $t$ departs from the focal time $t=0$. Figure~\ref{fig:skyrm} shows the intensity distribution and the polarization state distribution of one of these pulses in the $y=0$ plane, at two propagation instants, for $\boldsymbol{p}_{1,2}$ chosen as $(1,\mp\ui,0)$ and $\eta=0.76$. It can be observed that almost all full polarization states are represented within the shown region. 

\section{Concluding remarks}
We proposed a method for generating exact solutions of the scalar or (divergence-free) vector wave equations that represent STOVs, valid even for large divergence angles, and obtained the conditions under which these solutions are round at the focal time. Several generalizations to these formulas can be considered. First, while we focused on round STOVs, the conditions between the parameters $q$ and $s$ can be relaxed to obtain STOVs with different ellipticities. Also, the coefficients of the operator in Eq.~\eqref{eq:scalaroperator} can be changed and a term proportional to $\partial_y$ can be introduced to tilt the OAM to any arbitrary direction. Higher-order operators can also be considered that induce higher magnitudes of the OAM, although these higher-order vortices would be unstable,  disintegrating into simple vortices upon propagation. Finally, the vectorization operator in Eq.~\eqref{eq:electromagneticoperator} can be chosen differently (as long as it is divergence-free and commutes with the Laplacian), and/or the vector $\boldsymbol{p}$ can be chosen to include significant longitudinal components. 

This theoretical model was used to propose variants of STOVs whose transverse field components cover all possible states of paraxial polarization over sagittal planes that are normal to the direction of the OAM. Although we did not calculate it here, the Skyrme density of this polarization coverage does not change sign within the region in which the field is significant and for times near the focal time, so these pulses can be regarded as pulsed optical implementations of Skyrmions. 

\section*{Funding}

M.A.A. acknowledges funding from ANR-21-CE24-0014. R.G.C. acknowledges funding from the Labex WIFI 
(ANR-10-LABX-24, ANR-10-IDEX-0001-02 PSL*).

\section*{Acknowledgements}

The authors thank Konstantin Bliokh and Etienne Brasselet for useful discussions.

\section*{Disclosures}

The authors declare no conflicts of interest.

\section*{Data Availability Statement}

No data were generated or analyzed in the presented research.

\appendix

\section{Derivation of the equation for a complex-focus pulse with Poisson-like spectrum}
\label{sec:poisson}

The complex time-dependent field can be calculated as a superposition of monochromatic components according to
\begin{align}
E\left(\boldsymbol{r},t\right)=&\int_{-\infty}^{\infty}U(\boldsymbol{r},\omega)\, \exp{\left(-\ui \omega t\right)}d\omega,
\end{align}
where $U(\boldsymbol{r},\omega)$ satisfies the Helmholtz equation. By substituting the expressions for $U(\boldsymbol{r},\omega)$ in Eqs.~(\ref{eq:complexfield}) and (\ref{eq:Poissonspectrum}), we get
\begin{align}
E\left(\boldsymbol{r},t\right)=&\left(\frac{s}{\omega_0}\right)^{s+1}\frac{1}{\Gamma(s+1) R} \nonumber \\
&\times \int_{0}^{\infty} \omega^{s-1} \left\{\exp\left[{\left(-\frac{s}{\omega_0} -\ui t_-\right)}\omega\right] \right. \nonumber
\\
&\left.-\exp\left[{\left(-\frac{s}{\omega_0} -\ui t_+\right)}\omega\right]\right\}d\omega,
\end{align}
where $t_\pm=t\pm R/c-\ui q/c$. Note that $\omega^{s-1}$ can be substituted by $(\ui\partial/\partial t)^{s-1} $, which can be extracted from the integral. The remaining integrals are then straightforward to evaluate, as are the derivatives of the result, leading to
Eq.~(\ref{eq:Poissonfield}).

\section{Dimensions of the wave packet and condition to make it round}
\label{sec:blob}

We now carry out an asymptotic analysis to obtain the conditions under which the blob-like wave packet in Eq.~(\ref{eq:Poissonfield}) has a round shape at $t=0$.
The first step is to look at the focal point to see that one of the two terms in the expression dominates, which will simplify the derivation that will follow. Setting all arguments to zero we find
\begin{align}
    E\left(\boldsymbol{0},0\right)
     &=\frac{1}{-\ui \omega_0 q} \left[1-\frac{1}{\left(1+ \frac{2\omega_0 q}{s c}\right)^{s}}\right].
      \end{align}
We see that for large $s$ the first term (involving $t_-$) dominates. We then use in the derivations in this section the following approximation:
\begin{align}
    E\left(\bt{r},t\right)\approx \frac{1}{\omega_0 R\left(1+ \ui \frac{\omega_0}{s}t_-\right)^{s}}.
\label{eq:blob2}
\end{align}

Let us now study the extension of the wave packet in the transverse and longitudinal directions, to find conditions for the curvatures of the intensity to match. 
We start with the transverse profile. As $x$ and $y$ are interchangeable, we choose to work with the $x$-coordinate. Thus, setting $y$, $z$ and $t$ equal to zero, and ignoring the contribution involving $t_+$, the field is found to be approximately equal to
\begin{align}
    E\left(x \widehat{\bt x},0 \right)\approx\frac{1}{\omega_0 \sqrt{x^2-q^2}} \frac{1}{\left[1- \ui \frac{\omega_0}{cs}\left(\sqrt{x^2-q^2}+\ui q \right) \right]^{s}}.
\end{align}
 We study this expression in the focal region, so we assume that $x^2$ is much smaller than $q^2$ and we can approximate $\sqrt{x^2-q^2}$ as $-\ui q(1-x^2/2q^2)$. 
 The field's expression becomes
 \begin{align}
       E\left(x \widehat{\bt x},0\right)&\approx\frac{1}{-\ui \omega_0 q \left(1-\frac{x^2}{2 q^2}\right)} \frac{1}{\left( 1+\frac{\omega_0}{2csq}x^2\right)^{s}}
       \nonumber\\
  &\approx-\frac{1}{\ui \omega_0 q} \left(1+\frac{x^2}{2q^2}\right)\frac{1}{\left(1+\frac{\omega_0}{2 s c q} x^2\right)^{s}}.
\end{align}
The short-time-averaged intensity along the $x$-axis is approximately given by
\begin{align}
    I\left(x \widehat{\bt x},0\right)
    &=|E\left(x \widehat{\bt x},0\right)|^2\\
        &\approx\frac{1}{ \omega_0^2 q^2} \left(1+\frac{x^2}{2q^2}\right)^2\frac{1}{\left(1+\frac{\omega_0}{2s c q} x^2\right)^{2s}}
        \nonumber\\
    &\approx\frac{1}{\omega_0^2 q^2} \left(1+\frac{x^2}{q^2}\right)\left[1-(2s)\frac{\omega_0}{2 s c q} x^2\right]
    \nonumber\\
    &\approx\frac{1}{\omega_0^2 q^2} \left(1-C_{\perp}x^2\right)\nonumber\\
    &\approx\frac{\exp(-C_{\perp}x^2)}{\omega_0^2q^2},
    \label{eq:profx}
    \end{align}
where $C_{\perp}$ is the approximate relative curvature in the transverse direction of the pulse's envelope, given by
\begin{align}
    C_{\perp}=\frac{\omega_0}{cq}-\frac{1}{q^2}.
\end{align}

We now look at the longitudinal profile. Setting $x$, $y$ and $t$ to zero, the field is approximately equal to
       \begin{align}
       E\left(z \widehat{\bt z},0\right)&\approx
       \frac{1}{\omega_0 \left(z-\ui q \right)} \frac{1}{\left(1 - \frac{\ui \omega_0}{s} \frac{z}{c} \right)^{s}}.
        \end{align}
By assuming $|z|\ll q$, and keeping the terms up to second order in $z$, the short-time average intensity along the $z$-axis near the origin is seen to be given approximately by
      \begin{align}
      I\left(z \widehat{\bt z},0\right)&\approx|E\left(z \widehat{\bt z},0\right)|^2 \nonumber\\
       &\approx
      \frac{1}{\omega_0^2 q^2} \left(1-C_z z^2\right) \nonumber \\
      &\approx\frac{\exp(-C_z z^2)}{\omega_0^2 q^2} ,
      \label{eq:profz}
      \end{align}
where $C_z$ is the approximate relative curvature of the envelope of the field in the longitudinal direction, given by
\begin{align}
    C_z=\frac{1}{s}\frac{\omega_0^2}{c^2}+\frac{1}{q^2}.
\end{align}

To ensure the roundness of the wave packet, the curvatures along the transverse and longitudinal direction need to be equal, so that
\begin{align}
C_\perp = C_z \qquad \Leftrightarrow   \qquad   s= \frac{k_0^2q^2}{k_0q-2},
\end{align}
with $k_0=\omega_0/c$. Note that for the complex focus method to lead to solutions that resemble beams or pulses propagating towards larger $a$, one must make the assumption that $q$ is considerably larger than the wavelength for the central frequency $\omega_0$, since otherwise a non-negligible amount of counter-propagating plane wave components would be present. We therefore consider the case in which $k_0q$ is sufficiently large compared to unity, and we obtain the simple approximate result in Eq.~(\ref{eq:scalaroperator}),
where the second term is an order below the first and hence could be ignored, but we keep it as a first correction given its simplicity.


\section{Condition to obtain a round scalar STOV}
\label{sec:stov}

Let us propose a differential operator of the form
\begin{align}
    \widehat{W}=\frac{c}{\omega_0}\partial_x+\ui \left( \alpha \frac{1}{\omega_0}\partial_t+ \beta\frac{c}{\omega_0}\partial_z+ \ui \gamma\right).
\end{align}
We now find values for the parameters $\alpha$, $\beta$ and $\gamma$ that produce a round STOV when applying this operator to the pulse discussed in the previous two sections. 
For this, we apply this operator to the approximate form in Eq.~(\ref{eq:blob2}). Let us look at each derivative separately:
\begin{subequations}
\begin{align}
\partial_x
E\left(\boldsymbol{r},t\right)&\approx-\frac xR\left(
\frac1R+\ui\frac{\omega_0}{c}\frac{1}{1+\ui \frac{\omega_0}{s}t_-}\right) E\left(\boldsymbol{r},t\right),\\
\partial_z
E\left(\boldsymbol{r},t\right)&\approx-\frac{z-\ui q}R\left(
\frac1R+\ui\frac{\omega_0}{c}\frac{1}{1+\ui \frac{\omega_0}{s}t_-}\right) E\left(\boldsymbol{r},t\right),\\
\partial_t
E\left(\boldsymbol{r},t\right)&\approx-\ui\frac{\omega_0}{1+\ui \frac{\omega_0}{s}t_-}E\left(\boldsymbol{r},t\right).
\end{align}
\end{subequations}
where we used $
\partial_x R= x/R$ and $\partial_z R= (z - \ui q)/R$. 
For simplifying the derivation, we work with the quantity $\widehat{W}E/E$, which gives
\begin{align}
\frac{\widehat{W}E\left(\boldsymbol{r},t\right)}{E\left(\boldsymbol{r},t\right)} \approx& -\frac c{\omega_0}\frac{x}{R^2}+\ui\frac{ x}{R\left(1+\ui \frac{\omega_0}{s}t_-\right) }
+ \alpha \frac{1}{\left(1+\ui \frac{\omega_0}{s}t_-\right)}
\nonumber\\ & + \ui \beta \left[-\frac c{\omega_0}\frac{z-\ui q}{R^2}+\ui\frac{ (z- \ui q)}{R\left(1+\ui \frac{\omega_0}{s}t_-\right) }\right]
- \gamma.
\label{eq:WEonE}
\end{align}
Ideally, for $y=ct=0$ this factor should take the form of a vortex in the $xz$ plane centered at the origin. 
To simplify this expression, we now use expanded expressions of the factors $1 /(1+\ui \omega_0t_-/s)$, $1/R$ and $1/R^2$ up to second order in $x$ and $z$, assuming large values of $q$.
We fix $y=ct=0$, and 
to ensure that we use the appropriate root for the term we are using we write
$R= - \ui \sqrt{(q+\ui z)^2- x^2}$. 
The obtained expansions are as follows:
\begin{subequations}
\begin{align}
     \left(1+\ui \frac{\omega_0}{s}t_-\right)^{n}&\approx 1-\ui \frac{n \omega_0z}{cs}+n\omega_0\frac{csx^2+(1-n)q\omega_0z^2}{2c^2s^2q},\\
R^n&\approx(- \ui q)^n \left[1+ \ui n\frac zq - n\frac{x^2+(n-1)z^2}{2q^2}\right].
\end{align}
\end{subequations}
Using these results in Eq.~(\ref{eq:WEonE}) 
evaluated at $y=ct=0$, and keeping the terms up to second order in $x$ and $z$, we obtain:
\begin{align}
    \frac{\widehat{W}E\left(x \widehat{\bt x}+z \widehat{\bt z},0\right)}{E\left(x \widehat{\bt x}+z \widehat{\bt z},0\right)}
    \approx &K_0 + K_{x} x  +\ui K_{z} z +K_{xx} x^2\nonumber \\
    &+\ui K_{xz} xz + K_{zz} z^2 ,
    \label{eq:factor}
\end{align}
where $K_0$, $K_{x}$, $K_{z}$, $K_{xx}$, $K_{xz}$ and $K_{zz}$ are the following real constants:
\begin{subequations}
\begin{align}
    K_0&= \beta \frac{c}{q\omega_0} - \gamma + (\alpha-\beta),
    \label{eq:constant}\\
    K_{x}&=\frac{c - q \omega_0}{q^2 \omega_0},
    \label{eq:donut1}\\
    K_{z}&= -\frac{\beta c }{q^2 \omega_0} + \frac{(\alpha-\beta) \omega_0}{c s},
    \label{eq:donut2}\\
    K_{xx}&=-\frac{\beta}{2 q^2} + \frac{\beta c}{q^3 \omega_0} + (\beta-\alpha) \frac{\omega_0}{ 2 c s q} ,\\
    K_{xz}&=\frac{1}{q^2} - \frac{2 c}{q^3 \omega_0} - \frac{\omega_0}{c q s},\\
     K_{zz}&=\left(\beta-\alpha\right)\frac{\omega_0^2}{c^2 s^2}-  \frac{\beta c}{\omega_0 q^3}.
\end{align}
\end{subequations}

To solve for $\alpha$, $\beta$ and $\gamma$, we need to find three equations in terms of these five coefficients. Having an isotropic vortex at the origin imposes the first two conditions:
\begin{subequations}
\begin{align}
    K_0&=0,
    \label{eq:eone}\\
    K_{x}&= \pm K_{z}.
    \label{eq:etwo}
\end{align}
\end{subequations}
A nicely symmetric vortex would ideally also require that the remaining coefficients vanish. However, it is not possible to simultaneously satisfy the resulting five constraints. 
Therefore, to find the best possible third constraint, 
we first change to polar coordinates according to $x= r \sin{\varphi}$ and $z= r \cos{\varphi}$, so that Eq.~(\ref{eq:factor}) can then be rewritten as
\begin{multline}
    \frac{\widehat{W}E\left(x \widehat{\bt x}+z \widehat{\bt z},0\right)}{E\left(x \widehat{\bt x}+z \widehat{\bt z},0\right)}
    \approx  \ui K_{z} r \exp(\pm\ui \varphi)\\
    +r^2 \left[B+C_+\exp(2\ui\varphi)+C_-\exp(-2\ui\varphi)\right],
\end{multline}
where
\begin{subequations}
\begin{align}
    B&=\frac{K_{xx}+K_{zz}}2,\\
    C_{\pm}&=\frac{K_{zz} - K_{xx}\pm K_{xz}}4. 
    \end{align}
\end{subequations}

The modulus squared of this expression gives
\begin{multline}
   \Bigg |\frac{\widehat{W}E\left(x \widehat{\bt x}+z \widehat{\bt z},0\right)}{E\left(x \widehat{\bt x}+z \widehat{\bt z},0\right)} \Bigg | ^2 \approx  K_z^2r^2\\
    -2K_zr^3\Big[(C_{\pm}-B)\sin\varphi+C_{\mp}\sin3\varphi\Big ].
\end{multline}
The term that disrupts the most the roundness of the resulting intensity pattern is that proportional to $\sin\varphi$. 
Therefore, depending on the choice of vorticity, corresponding to the choice of sign in  Eq.~(\ref{eq:etwo}), we impose the condition $C_{\pm}=B$, which gives
\begin{align}
    3K_{xx}+K_{zz}=\pm K_{xz}.
    \label{eq:ethree}
\end{align}
Equations~(\ref{eq:eone}), (\ref{eq:etwo}) and (\ref{eq:ethree}) then form a system 
that 
can be solved to obtain the solutions for $\alpha$, $\beta$ and $\gamma$. 
For the top choice of the sign, these solutions are:
\begin{subequations}
    \begin{align}
        \alpha&=-\frac{5 c^3 s^2 + 3 c^2 q\omega_0 s (1-s) - c q^2 \omega_0^2(2-s)  + 4 q^3 \omega_0^3}{c q \omega_0 (c s - 2 q \omega_0 - 3 q s \omega_0)},\\
        \beta&=-\frac{c^2 s - 2 c q \omega_0 + c q s \omega_0 + 4 q^2 \omega_0^2}{c (c s - 2 q \omega_0 - 3 q s \omega_0)},\\
        \gamma&=\frac{c^2 s + 5 c^2 s^2 - 2 c q \omega_0 + 3 c q s \omega_0 - 3 c q s^2 \omega_0 + 4 q^2 \omega_0^2}{q \omega_0 (-c s + 2 q \omega_0 + 3 q s \omega_0)}.
    \end{align}
    \end{subequations}
which can be approximated, using $s\approx k_0 q$ by:
\begin{align}
    \alpha&\approx\frac{2}{3}-\frac{26}{9s},\qquad
    \beta\approx\frac{5}{3}-\frac{32}{9s},\qquad
    \gamma\approx-1+\frac{7}{3s}.
\end{align}

\section{Condition to obtain a round vector STOV}
\label{sec:elect}

To find the parameters that make a round vector STOV, we consider the short-time-averaged intensity
\begin{align}
    I(\boldsymbol{r},t) = ||\bf E_{\rm STOV}(\boldsymbol{r},t;\boldsymbol{p})||^2.
\end{align}
As was done in Appendix~\ref{sec:stov}, we evaluate the resulting expression at $ct=y=0$ and express $x$ and $z$ in polar coordinates. 
However, it turns out that a rounder short-time-averaged intensity profile is achieved if we allow for a small lateral displacement of magnitude $x_0$ for the intensity minimum away from the origin; we therefore write $x=x_0+r\sin\varphi$ and $z=r\cos\varphi$, and separate the result in terms of powers of $r$. 
We can then impose a series of constraints: 
\begin{enumerate}
    \item[(i)] that the intensity at the center (the part that is independent of $r$) be zero or as small as possible; 
\item[(ii)] that the part of the intensity linear in $r$ (which happens to also be proportional to $\sin\varphi$) be zero; 
\item[(iii)] that the coefficient of $r^2$ be independent of $\varphi$; and 
\item[(iv)] that the part of the coefficient of $r^3$ that oscillates as $\sin\varphi$ be zero (the remaining part oscillating as $\sin3\varphi$).
\end{enumerate}
The idea is to use these four constraints to determine $\alpha$, $\beta$, $\gamma$ and $x_0$.  
The expressions for these constraints are long and difficult to solve. To simplify the solution, we first consider the leading terms of the constraints in the limit of large $q$. By using computer algebra, we show that constraints (i), (ii) and (iv) all give the same relation to leading order:
\begin{subequations}
\begin{align}
    \beta+\gamma-\alpha={\cal O}(q^{-1}),
\end{align}
while constraint (iii) gives, once this prior constraint is taken into account,
\begin{align}
    \beta=\alpha\pm1+{\cal O}(q^{-1}),\qquad 
    \gamma=\mp1+{\cal O}(q^{-1}).
\end{align}
\end{subequations}
Let us choose the top choice for the sign (the other choice would just reverse the sign of the orbital angular momentum.) 
We then look at the next order corrections for $\alpha$, $\beta$, and $\gamma$, by looking at the corresponding corrections for the four constraints. 
It turns out that constraint (iii) has two parts, one independent of $\boldsymbol{p}$ and one that depends on this choice; the first part cannot be fully satisfied, so we only use the second in combination with the remaining constraints. 
We then find that the lateral displacement of the center is $x_0=-1/(7k_0)$, namely a small fraction of the main wavelength, and the coefficients are given by
\begin{subequations}
\begin{align}
    \alpha=&\frac67+\left(\kappa+\frac{237}{98}\right)\frac c{\omega_0 q},\\
    \beta=&\frac{13}7+\kappa\frac{c}{\omega_0 q},\\
    \gamma=&-1+\frac{237}{98}\frac{c}{\omega_0 q},
\end{align}
\end{subequations}
where the constant $\kappa$ is a correction that is not fixed by the constraints. Numerical tests for small $q$ show that the STOV remains fairly round if this free constant is chosen to eliminate the correction to $\alpha$. The coefficients are then, when written in terms of $s$
\begin{align}
    \alpha=\frac67,\qquad
    \beta=\frac{13}7-\frac{237}{98s},\qquad
    \gamma=-1+\frac{237}{98s}.
\end{align}
\section{Paraxial limit}
\label{sec:paraxial}

We now give the derivation for obtaining the limit expression of the pulse when approaching the paraxial regime. 
We start from the approximate expression in Eq.~(\ref{eq:blob2}), which we write as
\begin{align}
    E\left(\boldsymbol{r},t\right)\approx\frac{1}{\omega_0 R} \left(1+ \ui \frac{\omega_0}{s}t_-\right)^{-s},
    \label{eq:Poissonsfield}
\end{align}
where we recall that $R=\sqrt{\rho^2+(z-\ui q)^2}$ with $\rho^2=x^2+y^2$, and $t_-=t- R/c-\ui q/c$.

Since, in the paraxial regime, the pulse tends to a Gaussian, we aim at transforming Eq.~(\ref{eq:Poissonsfield}) into an expression comprising an exponential.
More specifically, we are looking for a correspondence between expansions up to second order of, on the one hand, an expression of the form $\left(1+ \epsilon\right)^{-N}$ and, on the other hand, an exponential of the form $\left[\exp\left(\epsilon + \varphi\right)\right]^{-N}$, where $|\epsilon|$ is a small quantity and $\varphi$ is introduced in the exponential as a possible correction term. We start by considering the expansion of both expressions up to second order:
\begin{subequations}
\begin{align}
\left(1+ \epsilon\right)^{-N}\approx 1 - N \epsilon + \frac{N (N+1)}{2}\epsilon^2,
 \label{eq:expand1}
\end{align}
and
\begin{align}
\left[\exp\left(\epsilon + \mu\right)\right]^{-N} &\approx 1 - N \left( \epsilon + \mu\right) + \frac{N^2}{2}\left(\epsilon + \mu\right)^2 \nonumber \\
&\approx 1 - N \epsilon - N \mu + \frac{N^2}{2} \epsilon^2,
\label{eq:expand2}
\end{align}
\end{subequations}
where we assumed that $|\mu|$ is much smaller than $|\epsilon|$, so we can neglect the terms in higher orders of $\mu$ or products of $\mu$ with $\epsilon$. 
Therefore, by equating  Eqs.~(\ref{eq:expand1}) and (\ref{eq:expand2}), we find that $\mu$ must be chosen as 
\begin{align}
\mu=-\frac{\epsilon^2}{2},
\end{align}
so that
\begin{align}
\left(1+ \epsilon\right)^{-N}\approx \exp\left(-N\epsilon +N \frac{\epsilon^2}{2}\right).
\end{align}
Using this result, we can write an approximation for Eq.~(\ref{eq:Poissonsfield}) valid for small $|\omega_0t_-/c|$ as
\begin{align}
     E\left(\boldsymbol{r},t\right)&\approx
     \frac{1}{ \omega_0 R} \exp\left(-\ui \omega_0t_- -\frac{\omega_0^2t_-^2}{2s} \right)
     \label{eq:Poissons2}.
\end{align}
The expansion of $R$ is the paraxial regime is given by:
\begin{align}
R \approx z-\ui q + \frac{\rho^2}{2(z-\ui q)}.
\end{align}
We can then write $t_-$ as
\begin{align}
t_- \approx t - \frac{z}{c}  - \frac{\rho^2}{2c\left(z-\ui q\right)}.
\end{align}
As is well known, while the paraxial approximation requires keeping terms of up to second order in the transverse variable in the exponent, a zeroth order approximation suffices for the amplitude, so we can approximate the prefactor $1/R$ as $1/(z-\ui q)$ in Eq.~\eqref{eq:Poissons2}.
Finally, using the condition for making the pulse round, $\omega_0/sc \approx 1/q$, we get 
the expression in Eq.~(\ref{eq:Gaussian_pulse}).

\bibliography{Bibliography}

\end{document}